# Anisotropic linear magnetoresistance in nanoflakes of Dirac semimetal $NiTe_2$

Ding Bang Zhou,[1,2] Kuang Hong Gao,[1,*] Tie Lin,[3] Yang Yang,[1] Meng Fan Zhao,[4] Zhi Yan Jia,[4] Xiao Xia Hu,[2] Qian Jin Guo,[2] and Zhi Qing Li[1]

[1]*Tianjin Key Laboratory of Low Dimensional Materials Physics and Preparing Technology, Department of Physics, Tianjin University, Tianjin 300354, China*

[2]*Analysis and Testing Center of Tianjin University, Tianjin 300072, China*

[3]*State Key Laboratory of Infrared Physics, Shanghai Institute of Technical Physics, Chinese Academy of Sciences, Shanghai 200083, China*

[4]*Institute of Quantum Materials and Devices, Tiangong University, Tianjin 300387, China*

## Abstract

This work investigates the magnetotransport properties of exfoliated $NiTe_2$ nanoflakes with varying thicknesses and disorder levels, unveiling two distinct physical mechanisms governing the observed anisotropic linear magnetoresistance (MR). For the perpendicular magnetic field configuration, the well-defined linear MR in high fields is unambiguously attributed to a classical origin. This conclusion is supported by the proportionalities between the MR slope and the carrier mobility, and between the crossover field and the inverse of mobility. In stark contrast, the linear MR under parallel magnetic fields exhibits a non-classical character. It shows a pronounced enhancement with decreasing flake thickness, which correlates with an increasing hole-to-electron concentration ratio. This distinctive thickness dependence suggests an origin in the nonlinear band effects near the Dirac point, likely driven by the shift of the Fermi level. Furthermore, the strengthening of MR anisotropy with enhanced interlayer transport contradicts the prediction of the guiding-center diffusion model for three-dimensional systems. Our findings highlight the critical roles of band topology and structural dimensionality in the anomalous magnetotransport of Dirac semimetals.

* Corresponding author, E-mail: khgao@tju.edu.cn

## I. INTRODUCTION

Compared to conventional quadratic magnetoresistance (MR) with only a few percent at high magnetic fields, nonsaturating linear MR (resistance linearly increases with external field) attracts much attention for application in magnetic sensors and memory devices. A prevailing mechanism to explain the linear MR behavior is a classical model in which disorder is proposed to give rise to a mixture of Hall and longitudinal resistances [1]. Apparently, the classical linear MR should be absent in a magnetic field applied parallel to the direction of excitation current. However, there appears a linear parallel field dependence of MR in some Dirac materials such as $WTe_2$, $Cd_3As_2$, $FeSe_{0.5}Te_{0.5}$, TlBiSe, and $Ag_{0.05}PdTe_2$ in recent experimental reports [2-6]. To explain the linear MR behavior in parallel field, a quantum model may be adopted [7], in which the linear MR emerges when only the lowest Landau level is occupied by carriers. But this model usually requires very low carrier concentration and high magnetic field, which limits its applicability. Since the anomalous linear MR appears specifically in Dirac materials, it is natural to speculate that the anomalous linear MR may correlate with the band topology. In order to test this speculation, we choose $NiTe_2$ as a material platform since it has been identified as one of the typical Dirac semimetal [8]. Because of its Dirac point very close to the Fermi level [9,10], $NiTe_2$ has been widely studied. Until now, it has revealed a range of interesting phenomena including spin-polarized surface state [11], Lifshitz transition [12], pressure-induced superconductivity [13]. Recently, a positive linear MR is reported when the field is perpendicular to sample surface [14,15], but its origin has not been fully explored.

In this work, we investigate the electrical transport properties of $NiTe_2$ nanoflakes with varying disorder. A well-defined linear MR is observed. The observed linear MR is found to be anisotropic. For the linear MR in perpendicular (magnetic field is perpendicular to the current direction) and in parallel (field is parallel to the current direction) external fields, there are two different origins: the former can be attributed to the disorder and the latter likely arises from the nonlinear effect of the energy bands.

## II. EXPERIMENTAL METHOD

$NiTe_2$ single crystal was fabricated by chemical vapor transport method. Powders of Ni (99.999%) and Te (99.999%) with a 1:2 molar ratio were sealed in a quartz tube under high vacuum. The quartz tube was heated from room temperature up slowly to 973 K in a furnace. Then the temperature was maintained for 4 days. Last, the quartz tube was slowly cooled to room temperature. The grown $NiTe_2$ single crystals have been characterized by transmission electron microscopy (TEM), X-ray diffraction (XRD), and energy dispersive X-ray spectroscopy (EDS) measurements. Through the mechanical exfoliation of the grown $NiTe_2$ single crystals, nanoflakes with different thicknesses were obtained. The exfoliated nanoflakes were transferred onto silicon substrates covered by a ~300 nm $SiO_2$ layer. The thicknesses of nanoflakes were determined by atomic force microscopy (Table I). A standard photolithography was used to define the measured patterns. For the ribbon-shaped nanoflakes, four Ohmic contacts were prepared by thermal evaporation of Ti/Au [Fig. 1(d)], while six Ohmic contacts were deposited for the wider nanoflakes [Fig. 1(e)]. Nine nanoflakes studied in this work are referred to as samples #1-#9, the thickness of which increases from 56.4 to 219.0 nm in sequence (Table I). The magnetotransport measurement were performed in a physical property measurement system (Quantum Design).

## III. RESULTS AND DISCUSSION

Figure 1(a) shows XRD pattern of the grown $NiTe_2$ single crystal. One can see that four peaks correspond to $NiTe_2$ structure, which can be indexed as (00*L*) reflections. This indicates that the exposed surface is the crystallographic *ab* plane (i.e., normal to the *c* axis). According to the position of the (002) peak, we obtain lattice parameter *c* ($= 0.5260\ \mathrm{nm}$), in accordance with the reported value of the stoichiometric $NiTe_2$ [16]. Inset of Fig. 1(a) is a high-resolution TEM image of the grown single crystal. A well-defined atomic ordered arrangement is observed, without secondary phase. The EDS images of two elements Ni and Te, as shown respectively in Fig. 1(c) and 1(d), exhibit uniform distributions. The atomic ratio of Ni:Te is determined to be 1:1.9 that is consistent with 1:2 of the stoichiometric $NiTe_2$ within experimental error.

Figure 2(a) shows the longitudinal resistivity $\rho_{xx}$ as a function of temperature *T*

for all studied samples. One can see that $\rho_{xx}$ decreases on reducing $T$ from 300 K down to ~20 K for any given sample, indicating a metallic behavior. In the lower $T$ range of 2- ~20 K, $\rho_{xx}$ tends to saturate with a residual resistivity. We can define the residual resistivity ratio (RRR) as $\rho_{xx}(300\ \mathrm{K}) / \rho_{xx}(2\ \mathrm{K})$ that varies between 13 and 23 for our samples (Table I). These values are far smaller than the reported results in the literature [8,14,15], indicating a strong disorder in our $NiTe_2$ nanoflakes. It is worthy noting that RRR does not correlate with the thickness of nanoflakes. When the out-of-plane magnetic field $B$ (i.e., perpendicular to the current) is applied, as shown in Fig. 2(b) for sample #6 as an example, one can find that $\rho_{xx}$ progressively increase as $B$ increases in the $T$ range of 2-~100 K at a given temperature. This is suggestive of the appearance of a positive MR at low temperatures. Figure 2(c) shows the MR curves in perpendicular fields at selective temperatures for sample #6. Here, MR is defined as $\mathrm{MR} = [\rho_{xx}(B) - \rho_{xx}(B=0)] / \rho_{xx}(B=0) \times 100\%$ . The positive MR are observed for all selective temperatures. At any given temperature, a well-defined linear MR appears in high fields while a parabolic MR is observed near zero field ($|B|$ is lower than ~2 T). And the linear MR at high fields is gradually suppressed on increasing temperature. For other samples, as shown in Fig. 2(d), the well-developed linear MR behavior is also observed in high fields.

In Dirac materials with linear dispersion, Abrikosov proposed the quantum model to explain the linear MR [7]. It is noteworthy that the appearance of quantum linear MR should fulfill equality $N \ll (eB/\hbar)^{3/2}$, where $N$ is carrier concentration and $\hbar$ is the reduced Planck's constant [7]. This means that one should observe quantum linear MR when $B \gg (\hbar/e)N^{2/3}$. In order to check this condition, Hall measurements were performed in samples #2 and #6-#8. Figure 3(a) shows the Hall resistivity $\rho_{xy}$ as a function of $B$ at various temperatures for a representative sample #6. The slope of these curves is positive at any given temperature, indicating that transport behavior is dominated by holes. Importantly, these curves are nonlinear, implying a multi-carriers transport behavior. To extract carrier concentrations and mobilities, a two-band model [17] is adopted to fit the experimental data. In this model, $\rho_{xy}$ is given as

$$\rho_{xy}(B)=\frac{(n_{\rm p}\mu_{\rm p}^2-n_{\rm e}\mu_{\rm e}^2)B+(n_{\rm p}-n_{\rm e})\mu_{\rm p}^2\mu_{\rm e}^2B^3}{e[(n_{\rm e}\mu_{\rm e}+n_{\rm p}\mu_{\rm p})^2+(n_{\rm p}-n_{\rm e})^2\mu_{\rm e}^2\mu_{\rm p}^2B^2]}, \quad (1)$$

where $n_{\rm e}$, $n_{\rm p}$, $\mu_{\rm e}$ and $\mu_{\rm p}$ are electron and hole concentrations and mobilities, respectively. As seen in Fig. 3(a), the experimental data can be well reproduced by Eq. (1). The fit-obtained $n_{\rm e}$ and $n_{\rm p}$ , as seen in Table I, are in the order of $10^{21}$-$10^{22}$ $cm^{-3}$. This indicates that the value for $(\hbar/e)N^{2/3}$ ($N=n_{\rm e},n_{\rm p}$) is larger than 658 T. The value is far higher than the maximum field of 9 T we can realize. That is, the equality $B\gg(\hbar/e)N^{2/3}$ is not satisfied in our samples. The observed linear MR therefore cannot be explained by the quantum model. Furthermore, the observed MR, as seen in Fig. 2(c) and 2(d), does not show quantum oscillations up to 9 T. This means that there is no Landau quantization due to large broadening of Landau level possibly resulting from the strong disorder (corresponding to low carrier mobility). This further excludes the quantum model.

The fit-obtained $\mu_{\rm e}$ and $\mu_{\rm p}$, as seen in Table I, vary between 98 and 2821 $cm^2V^{-1}s^{-1}$. These values are lower than that in some typical Dirac materials like $Cd_3As_2$ and NbP [18,19] but comparable to that in earlier reports where the band topology of $NiTe_2$ persists [8,15]. Although low mobility is suggestive of a strong disorder, it is not indicative of a broken band topology [20] because the broken band topology usually induces an insulating state or Anderson localization [21,22] that does not appear in our samples.

In addition to the quantum model, the “hot spots” theory is also proposed to explain the linear MR [23], According to this theory, the scattering from the charge density wave fluctuations in the hot spots of Fermi surface can give rise to the linear MR. Because there is no evidence of the presence of charge density wave state in $NiTe_2$, this mechanism may be invalid for our samples. Furthermore, the Berry curvature is proposed to induce a large linear MR in Weyl semimetal, and this linear MR is expected to be larger when the Fermi level is more near to the Dirac point [24]. However, we note that the first derivative of linear MR [i.e., $d(\mathrm{MR})/dB$ ] does not exhibit a distinct thickness dependence at 2 K, as shown in the inset of Fig. 3(b). Meanwhile, the Fermi level is gradually away from the Dirac point on decreasing thickness (shown later) for our samples at 2 K. This implies that the magnitude of the observed linear MR is not related to the distance between the Fermi level and Dirac point, inconsistent with the theory of the Berry curvature-induced linear MR.

Alternatively, we find that the classical model can provide a reasonable explanation because of two reasons. First, as seen in the left inset of Fig. 3(a), $d(\mathrm{MR})/dB$ in high fields is proportional to the effective Hall mobility [ $\mu^* = (n_\mathrm{e}\mu_\mathrm{e} + n_\mathrm{p}\mu_\mathrm{p})/(n_\mathrm{e} + n_\mathrm{p})$ ]. That is, $d(\mathrm{MR})/dB \propto \mu^*$, consistent with the prediction of the classical model. Second, an empirical expression $\mathrm{MR} = (k^2B^2 + a^2)^{0.5} - a$ ($k$ and $a$ are two parameters) is usually used to describe MR including both parabolic and linear field dependences [25-27]. We use it to fit the MR curves. As seen in Fig. 2(c) and 2(d) for some examples, the experimental data can be well described by this expression (dashed lines in the negative $B$ region are fitting curves). According to two fit-obtained parameters $k$ and $a$, one can calculate the crossover field $B_\mathrm{C} = a/k$ that indicates a transition from quadratic to linear $B$ dependence [25-27]. The calculated $B_\mathrm{C}$, as seen in the right inset of Fig. 3(a), is proportional to the inverse of effective Hall mobility $(\mu^*)^{-1}$, in agreement with the classical model. This, combined with the relation $d(\mathrm{MR})/dB \propto \mu^*$, supports that the observed linear MR is attributed to the classical model. The point is further confirmed by the RRR dependence of $d(\mathrm{MR})/dB$. The relation $d(\mathrm{MR})/dB \propto \mu^*$ also should be valid for different samples at a given temperature. Because Hall measurement cannot be performed in some our samples like #1, we use RRR instead of mobility to check the relation. Figure 3(b) shows $d(\mathrm{MR})/dB$ as a function of RRR for all studied samples at 2 K. One can see that $d(\mathrm{MR})/dB$ scales linearly with RRR, further supporting the classical model. Obviously, the aforementioned observation that $d(\mathrm{MR})/dB$ does not have a distinct thickness dependence [see inset of Fig. 3(b)] is due to the fact that RRR does not correlate with the thickness of nanoflakes (Table I).

Figure 4(a) shows the MR curves at 2 K in different tilted fields for sample #3. Here, $\theta$ is an angle between field and the normal direction of nanoflake [see the inset of Fig. 4(b)]. One can find that the linear MR in high fields is gradually suppressed on increasing $\theta$ from 0° to 90°. That is, the observed linear MR is anisotropic. Interestingly, even for $\theta = 90^{\circ}$, a well-developed linear $MR_{//}$ is still observed (hereafter, MR for $\theta = 90^{\circ}$ is defined as $MR_{//}$ for convenience). A similar phenomenon also appears in other samples, as seen in Fig. 4(b) for sample #8 as an example. Figure 4(c) shows $\rho_{xx}$ as a function of $\theta$ at 2 K with some selective fields for sample #3. One can see that $\rho_{xx}$ peaks at both $\theta = 0^{\circ}$ and $\theta = 180^{\circ}$ for a given field, but the maximum shows a slight difference at these two angles. The small difference (~9%) may arise from a residual misalignment of ~2–3 ° between the magnetic field plane and the sample *ab* plane, which needs further investigation. Meanwhile, $\rho_{xx}$ forms two valleys at both $\theta = 90^{\circ}$ and $\theta = 270^{\circ}$. Importantly, $\rho_{xx}$ values at these two valleys increase linearly with $B$, confirming the presence of linear $MR_{//}$ for $B$ applied parallel to the direction of current. Figure 4(d) shows the $MR_{//}$ curves at selective temperatures. The linear $MR_{//}$ in high fields is also gradually suppressed on increasing temperature, which is similar to the observation in Fig. 2(c) for $\theta = 0^{\circ}$. Despite there exists a similar temperature dependence, the observed linear $MR_{//}$ cannot be simply regarded as a partial contribution of that for $\theta = 0^{\circ}$. The inset of Fig. 4(a) shows the MR curves as a function of the perpendicular field component (i.e., $B\cos\theta$) for various tilted fields. As seen, all the curves are not collapsed onto a single curve, indicating that the linear $MR_{//}$ in parallel fields is not a component of that in the perpendicular field.

What causes the linear $MR_{//}$? Owing to the distinctive layered structure with expectedly weak interlayer coupling, the transport property of $NiTe_2$ should be two-dimensional electronically. In conventional two-dimensional systems like GaAs-base and Si-base heterostructures, both Zeeman effect and magnetic-orbit coupling may cause large positive $MR_{//}$ [28,29]. But the $MR_{//}$ arising from the two effects does not linearly depend on field. We therefore think that the observed linear $MR_{//}$ in our samples cannot be attributed to these two effects.

In semimetal with electron-hole compensation, a linear $MR_{//}$ possibly appears at charge neutrality with equal concentrations of electron and hole [30,31]. Considering that the ratio ($n_{\mathrm{p}} / n_{\mathrm{e}}$) of hole and electron concentrations is larger than unit for our samples (see $n_{\mathrm{p}}$ and $n_{\mathrm{e}}$ Table I), we thus insist that the electron-hole compensation cannot be used to explain the observed linear $\mathrm{MR}_{//}$, either. In addition, the linear $\mathrm{MR}_{//}$ originating from the aforementioned hot spots of the Fermi surface and Berry curvature [23,24] can be excluded either because both these two mechanisms are closely tied to the Lorentz force that is expected to be absent for $\mathrm{MR}_{//}$.

Recently, Song *et al*. proposed a guiding center diffusion model in which the linear MR comes from the trajectories of carriers influenced by a weak disorder [32]. It is noteworthy that this model is only applied to the three-dimensional system, and has been used to explain the linear MR in Dirac semimetal $Cd_3As_2$ [33]. If this model was valid for layered $NiTe_2$, the interlayer transport must play an important part. In order to clarify its influence on the linear $\mathrm{MR}_{//}$, $\rho_{xx}$ in titled field is further analyzed by using a three-dimensional scaling approach [34]

$$\rho_{xx}(B,\theta) = \rho_{xx}(\varepsilon_{\theta} B) \tag{2}$$

where $\varepsilon_{\theta}$ is a scaling factor. As shown in the inset of Fig. 5(a) for the same data of sample #8 in Fig. 4(b), the field scaling makes the MR curves at various angles be collapsed onto a single curve. This analysis allows us to extract $\varepsilon_{\theta}$. The main panel of Fig. 5(a) shows the extracted $\varepsilon_{\theta}$ as a function of $\theta$ for two representative samples #3 and #8. One can see that $\varepsilon_{\theta}$ vs $\theta$ curves can be well fitted by

$$\varepsilon_{\theta} = (\cos^2\theta + \gamma^{-2}\sin^2\theta)^{1/2}, \tag{3}$$

where $\gamma^2$ is a ratio of carrier effective mass for $\theta = 0^{\circ}$ and $90^{\circ}$. The parameter $\gamma$ reflects the anisotropy of carrier effective mass but cannot distinguish the specific properties of electrons and holes [34]. From the fits, $\gamma$ can be obtained. The fit-obtained $\gamma$ is given in Table I. It should be emphasized that we also extract $\gamma$ from the analysis of MR data with titled angle $\theta$ in the range of 180-270$^{\circ}$. As seen in Fig. 5(b), the extracted $\gamma$ (open circles) are comparable to that (closed triangles) obtained from the analysis with $\theta$ in the range of 0-90$^{\circ}$ for each sample. This hints that the slight difference of amplitude between $0^{\circ}$ and $180^{\circ}$ [see Fig. 4(c)] has a negligible influence on three-dimensional scaling analysis. Importantly, the extracted $\gamma$ from these two methods consistently increases by a factor of ~2 with thickness increasing from 56.4 nm to 219.0 nm. Such systematic trend reveals a gradual change of interlayer transport that has never been reported in bulk $NiTe_2$. The increased $\gamma$ implies that the interlayer transport are gradually enhanced with increasing thickness, consistent with the reported results in other layered compounds [35,36]. Apparently, the enhanced interlayer transport was expected to suppress the transport anisotropy if the three-dimensional guiding center diffusion model was valid. We then define the anisotropy as $\mathrm{MR/MR}_{//}$ [i.e., $\mathrm{MR}(\theta=0^{\circ})/\mathrm{MR}(\theta=90^{\circ})$. Unexpectedly, as seen in Fig. 5(c), $\mathrm{MR/MR}_{//}$ at 2 K and 9 T increases with $\gamma$ (here, $\gamma$ obtained from the MR analysis with $\theta$ in the range of 0-90$^{\circ}$ is used). That is, the transport anisotropy becomes more stronger with enhancing interlayer transport. This indicates that the guiding center diffusion model cannot be used to explain the observed linear $MR_{//}$ in our samples.

Furthermore, this model requires that the carrier mobility $\mu \gg 2\pi/B$ [37]. This condition tells us that the linear MR should appear when $B \gg 2\pi/\mu$. We find that it is not satisfied in our samples. Taking sample #6 as an example, we substituted $\mu^* = 1395.4$ $\mathrm{cm^2V^{-1}s^{-1}}$ at 2 K into the above inequality, and found that the linear MR should appear for $B \gg 45$ T according to the guiding center diffusion model. This is inconsistent with our observations that the linear MR occurs below 9 T for all our samples. This confirms that the observed linear $MR_{//}$ cannot be attributed to the guiding center diffusion model.

Importantly, we find that the first derivative of linear $MR_{//}$ (i.e., $\mathrm{dMR}_{//}/\mathrm{d}B$) exhibits a distinct thickness dependence. As seen in Fig. 5(d), $\mathrm{dMR}_{//}/\mathrm{d}B$ shows a linear increase by a factor of ~5 with decreasing thickness from 219.0 nm to 56.4 nm. The remarkable increase is completely different from the thickness dependence of linear MR in perpendicular fields [see the inset of Fig. 3(b)]. One therefore expects a non-classical origin for the linear $MR_{//}$.

Theoretically, a negative $MR_{//}$ was initially predicted due to the chiral anomaly in Dirac semimetals [38-40]. When a strong intervalley scattering is considered, the sign of $MR_{//}$ can be changed from negative to positive in nonquantizing magnetic fields [41,42]. Recently, it is proposed that, even without intervalley scattering, the nonlinear effect (bands are no longer linear away from the Dirac point) also assists driving the semimetal to show positive $MR_{//}$ [43]. When the Fermi level is farther away from the Dirac point, this effect is expected to become more important [43]. As a consequence, the nonlinear effect-induced positive $MR_{//}$ also becomes larger. If this effect causes the observed linear $MR_{//}$ for our $NiTe_2$ nanoflakes, the Fermi level must be gradually away from the Dirac point on decreasing thickness because of the enhanced linear $MR_{//}$ [see Fig. 5(d)]. Indeed, the ratio $n_p / n_e$ varies between 1.1 and 6.7, which shows a linear increase on decreasing thickness [see the inset of Fig. 5(d)]. This indicates that the observed linear $MR_{//}$ likely originates from the nonlinear effect. The observation that linear $MR_{//}$ is gradually diminished with temperature [Fig. 4(d)] may be due to phonon scattering, which awaits further study.

## IV. CONCLUSION

We study the magnetotransport properties of exfoliated $NiTe_2$ nanoflakes with different thicknesses. In perpendicular fields, a well-developed linear MR is observed. Through Hall data analysis, we found that the observed linear MR cannot be attributed to the quantum model. On the contrary, the classical model offers a reasonable explanation because $d(\mathrm{MR})/dB \propto \mu^*$ and $B_C \propto (\mu^*)^{-1}$. Most importantly, a well-defined linear $MR_{//}$ in parallel fields is observed. The observed linear $MR_{//}$ is enhanced with decreasing the thickness of nanoflakes. We think that it has a non-classical origin, and likely arises from the nonlinear effect of the band structure.

## ACKNOWLEDGMENTS

This work was supported by the National Natural Science Foundation of China (Grant Nos. 12174282 and 61974153) and the Frontier Fundamental Research Program of Tianjin University (Grant No. 2025XJ21-0005).

## DATA AVAILABILITY

The data that support the findings of this article are not publicly available. The data are available from the contact authors upon reasonable request.

**Figure captions**

**Figure 1** (a) X-ray diffraction pattern of $NiTe_2$ single crystal. Inset gives high-resolution transmission electron microscopy image. (b) and (c) are energy dispersive X-ray spectroscopy images of two elements Ni and Te. (d) and (e) are optical microscope images of samples #1 and #8.

**Figure 2** (a) Resistivity $\rho_{xx}$ as a function of temperature $T$ for all samples. (b) $\rho_{xx}$ vs $T$ plots at various magnetic fields $B$s for sample #6. (c) Magnetoresistance (MR) at some selective $B$s for #6. Dashed lines in the $B$ range of -9–0 T are fits according to the expression $\mathrm{MR} = (k^2B^2 + a^2)^{0.5} - a$. (d) MR at 2 K for some selective samples. Dashed lines in the $B$ range of -9–0 T are fits.

**Figure 3** (a) Hall resistivity $\rho_{xy}$ as a function of $B$ at various temperatures for sample #6. Dashed lines are fits according to Eq. (3). Left inset is the first derivative of linear MR (i.e., dMR/d$B$) in high fields as a function of effective mobility $\mu^*$. Dashed line provides a guide for the eye. Right inset shows crossover field $B_C$ as a function of the inverse of $\mu^*$. (b) dMR/d$B$ as a function of RRR at 2 K. Dashed line also provides a guide for the eye. Inset gives dMR/d$B$ as a function of thickness at 2 K.

**Figure 4** (a) MR at 2 K for sample #3 when $B$ is tilted along various directions. Inset shows the MR vs $B\cos\theta$ plots. (b) MR at 2 K for sample #8 when $B$ is tilted along various directions. Inset is a sketch of experimental configuration ($\theta$ is tilted angle). (c) $\rho_{xx}$ vs $\theta$ plots at 2 K at various $B$s for sample #3. (d) MR at selective $T$s for sample #3 when $\theta = 90^\circ$.

**Figure 5** (a) $\varepsilon_\theta$ vs $\theta$ at 2 K for samples #3 and #8. Solid lines are fits according to Eq. (3). Inset shows MR vs $\varepsilon_\theta B$ plots at various $\theta$s for sample #8. (b) The extracted parameter $\gamma$ as a function of the thickness. Closed triangles (open circles) are the data obtained from the analysis of MR data with titled angle $\theta$ in the range of 0-90$^\circ$ (180-270$^\circ$). Dashed line provides a guide for the eye. (c) MR/MR ($\theta$ = 90$^\circ$) as a function of $\gamma$. (d) dMR ($\theta$ = 90$^\circ$) /d$B$ as a function of thickness at 2 K. Inset is a ratio of the extracted hole concentration $n_p$ and electron concentration $n_e$ as a function of thickness at 2 K.

**Table I.** Electrical parameters of the exfoliated nanoflakes. RRR is residual resistivity ratio. $\gamma$ is the parameter obtained from the analysis of MR data with titled angle $\theta$ in the range of 0-90°. $n_e$, $n_p$, $\mu_p$ and $\mu_e$ are respectively fit-obtained hole and electron concentrations and mobilities according to Eq. (1) at 2 K.

| Sample | Thickness (nm) | RRR | $\gamma$ | $n_p$ ($\times 10^{22}$ cm$^{-3}$) | $n_e$ ($\times 10^{22}$ cm$^{-3}$) | $\mu_p$ (cm$^2$ V$^{-1}$s$^{-1}$) | $\mu_e$ (cm$^2$ V$^{-1}$s$^{-1}$) |
|---|---|---|---|---|---|---|---|
| #1 | 56.4 | 17.4 | 1.77 | - | - | - | - |
| #2 | 64.6 | 21.4 | 1.44 | 5.19 | 0.77 | 98.71 | 138.86 |
| #3 | 95.5 | 19.6 | 1.90 | - | - | - | - |
| #4 | 101.0 | 14.6 | 1.75 | - | - | - | - |
| #5 | 153.7 | 13.5 | 2.68 | - | - | - | - |
| #6 | 165.6 | 15.9 | 2.05 | 2.33 | 1.28 | 1324.09 | 1525.06 |
| #7 | 182.7 | 14.5 | 2.72 | 3.63 | 1.52 | 1090.31 | 1502.53 |
| #8 | 202.8 | 22.4 | 2.48 | 1.06 | 0.96 | 2820.42 | 2447.81 |
| #9 | 219.0 | 16.9 | 2.87 | - | - | - | - |

Figure 1

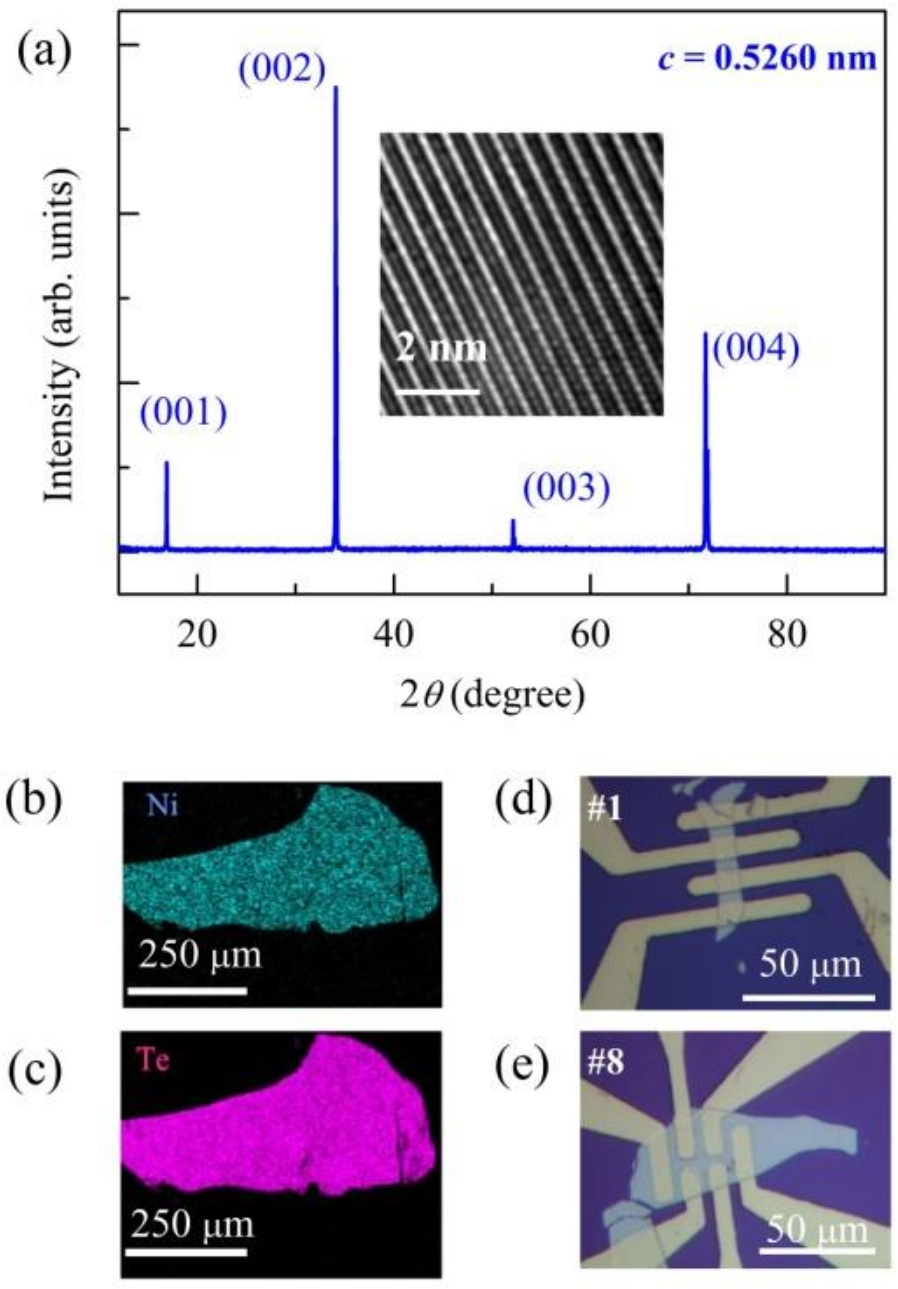

Figure 2

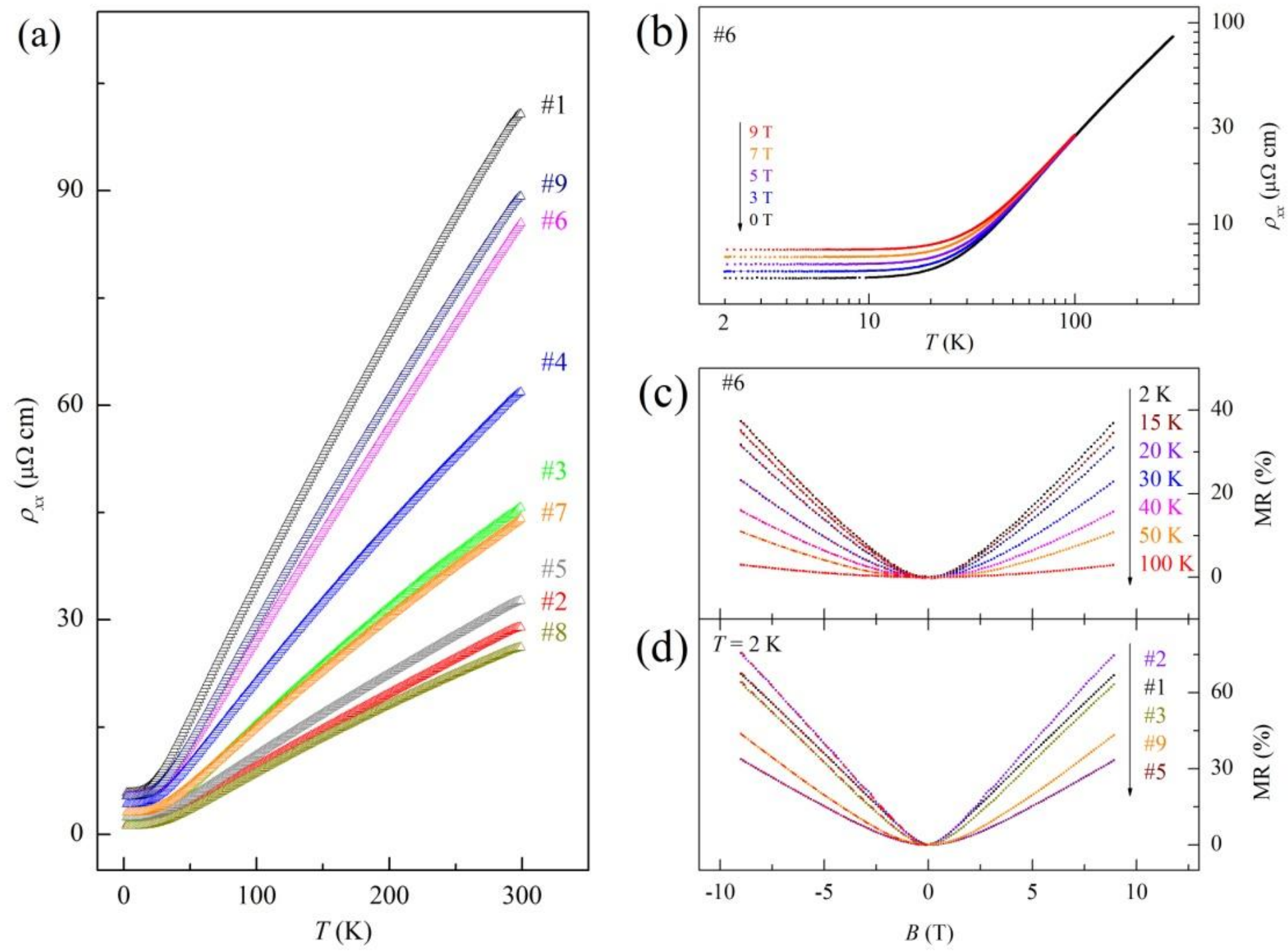

Figure 3

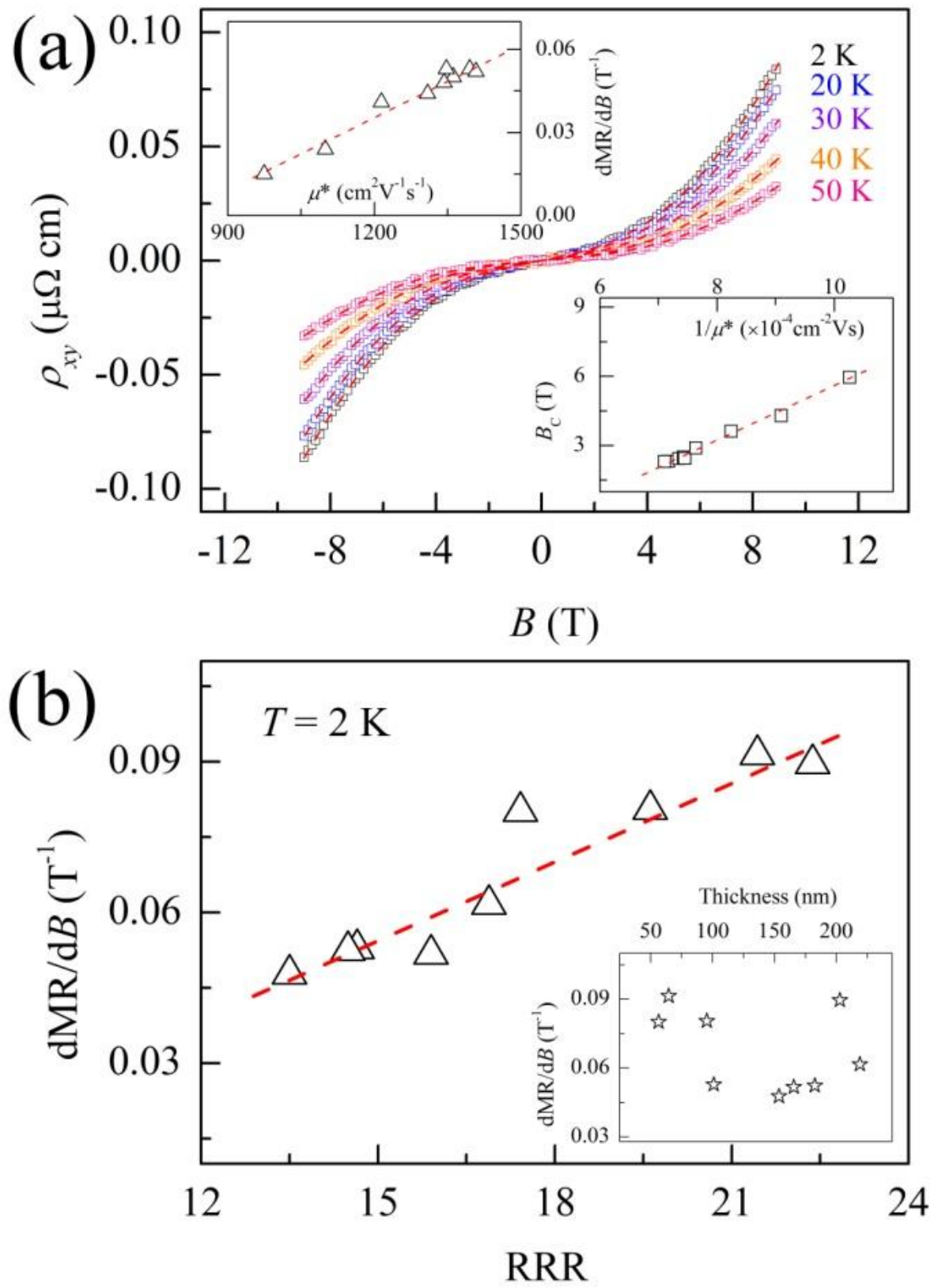

Figure 4

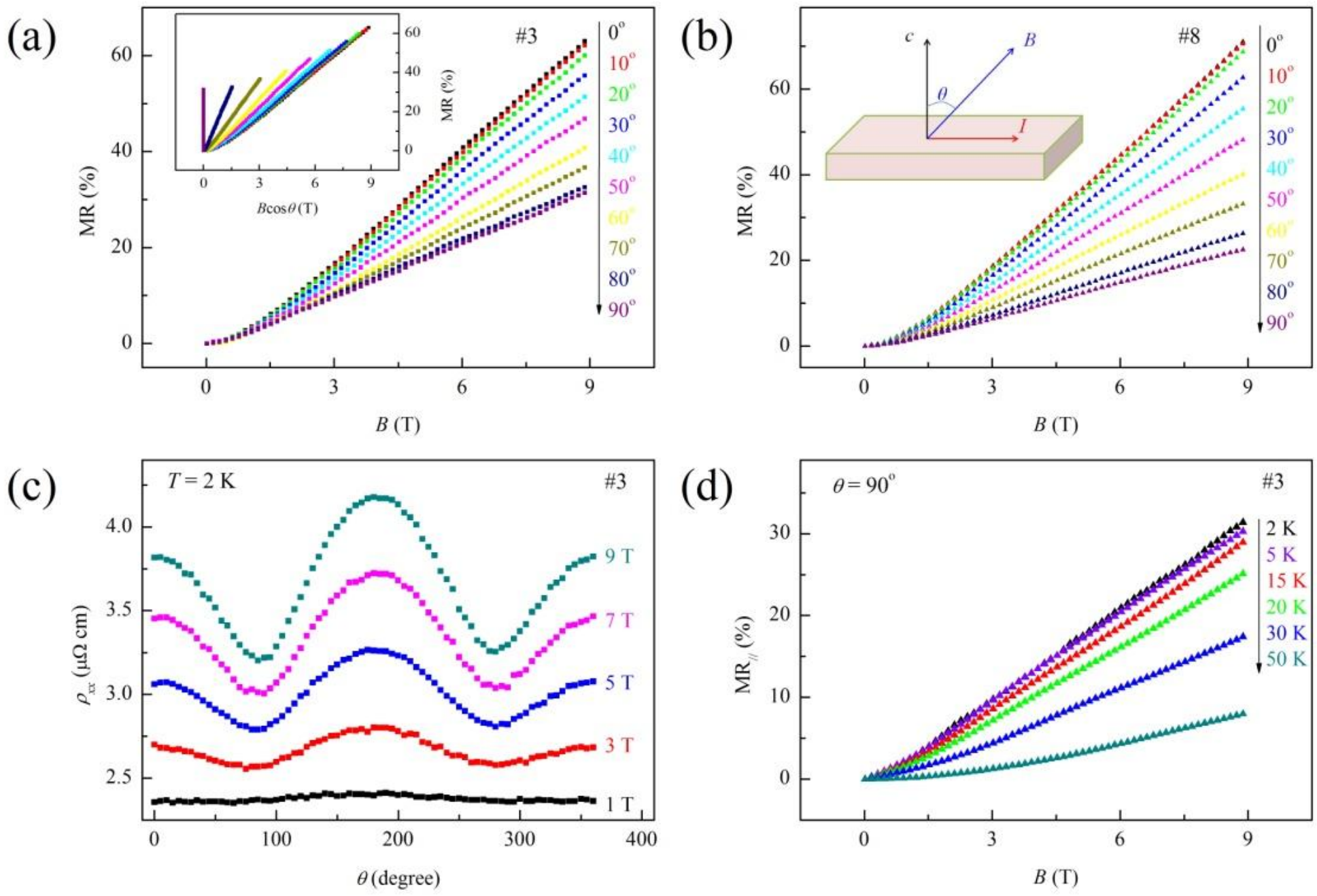

Figure 5

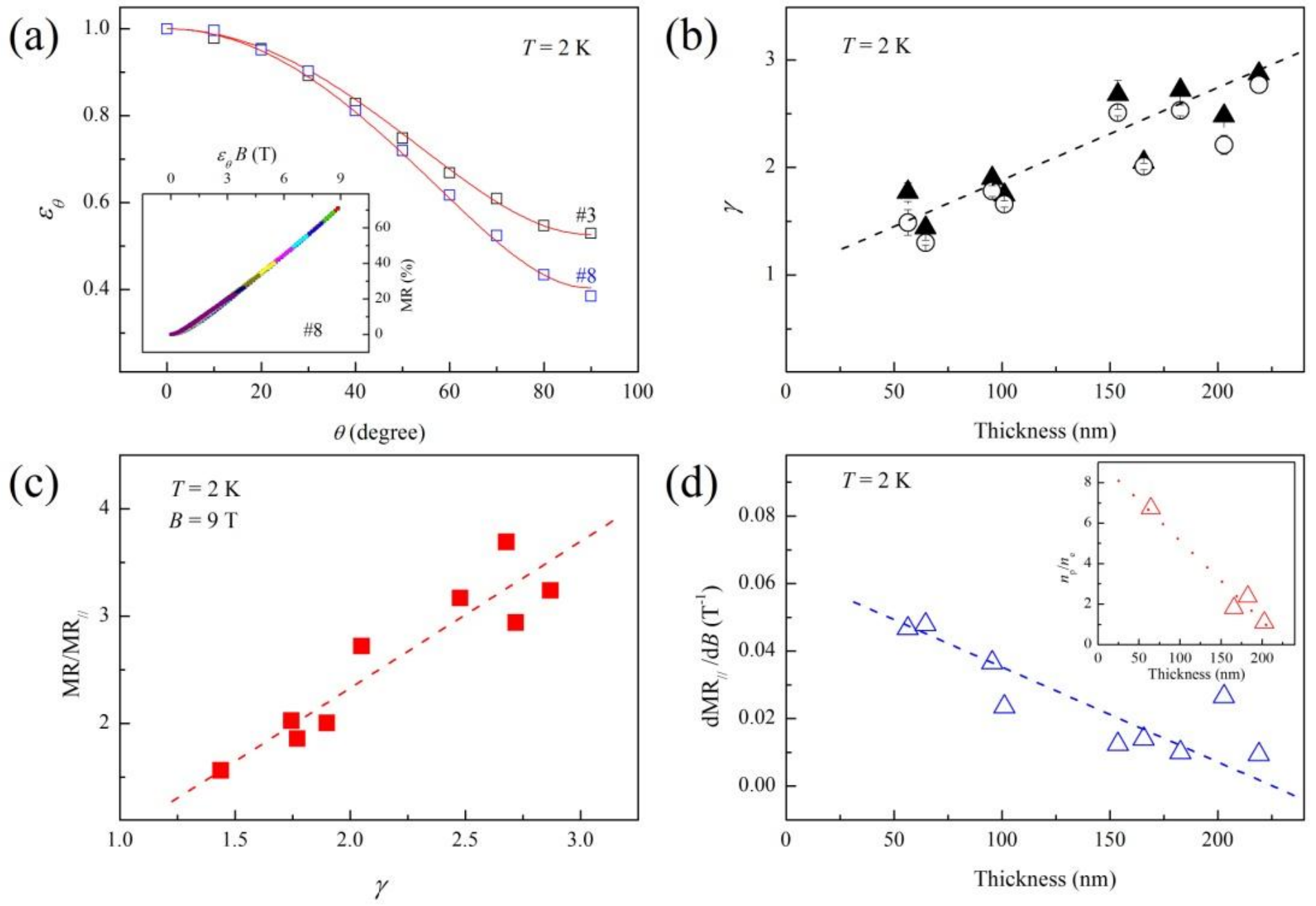